\def\vphi{\langle \phi \rangle}
\def\tphi{ \tilde\phi}
\def\gm{ \gamma}
\def\k{\kappa}
\def\mphi{m_{\phi}}

\def\l{\lambda}
\def\mt{m_t}
\centerline{\bf Some low energy effects of a light stabilized radion}
\centerline{\bf in the Randall-Sundrum model}
\vskip 1 true in
\centerline{\bf Uma Mahanta, Mehta Research Institute}
\centerline{\bf Chhatnag Road, Jhusi, Allahabad-211019, India}
\centerline{\bf and}
\centerline{\bf Subhendu Rakshit, Department of Physics}
\centerline{\bf University of Calcutta, Calcutta-700009, India}

\vskip .4 true in

\centerline{\bf Abstract}

In this paper we study some of the low energy effects of a light 
stabilized radion field in the Randall-Sundrum scenario. 
We find that the NLC 500 with its projected precision level
will be able to probe the radion contribution to $\k_v$
and $\l_v$ for values of $\vphi $ up to 500 Gev. On the other hand
the BNL E821 experiment will be able to test the radion contribution
to $a_{\mu}$ for $\vphi $= 1 Tev and $\mphi\le m_{\mu}$.
We have also shown that the higgs-radion mixing 
induces a 2.6\% correction in
the WWh coupling.
Finally  by comparing the 
radionstralung process with the higgsstralung process we have found
 that the
 LEPI bound of 60 Gev on the higgs mass based on 
$Z\rightarrow hl\bar{l}$  decay mode suggests a lower bound of about 35 Gev 
on the radion mass.


\vfill\eject

Recently Goldberger and Wise [1] have shown that the separation between the 
branes in the Randall-Sundrum (R-S) scenario [2] can be stabilised by a bulk
scalar field. They also showed [3]
 that if the large value of $kr_c\approx 12$
that is necessary for solving the heirarchy problem
arises from a small bulk scalar mass term then the modulus potential is 
nearly flat near its minimum. The R-S scenario therefore predicts
a modulus field which is much lighter than the Kaluza-Klein excitations
of the bulk fields which usually lie in the Tev range.
 The modulus field is therefore expected to be the first
experimental signal of R-S model. Its couplings to ordinary fields
on the visible brane is suppressed by the Tev scale and completely 
determined by general covariance. Although the radion vev is more or less
well determined in the R-S scenario its mass is an almost free
parameter. It turns out to be much lighter than 1 Tev only if
the mass of the bulk scalar field that stabilizes the modulus
is smaller than 1 Tev. It is the purpose
of this paper to investigate what bounds can be derived on the radion 
vev and particularly its 
mass from precision tests and direct searches at $e^+e^-$ collider.

The couplings of the modulus field $\phi$ to ordinary matter on the 
visible brane is given by [3]
$$L_I={T^{\mu}_{\mu}\over \vphi}\tphi .\eqno(1)$$

where $\phi$ is expanded about its vev as $\phi = \tphi + \vphi $.
In the context of the SM, ignoring the contribution of the higgs sector,
we have
 $T^{\mu}_{\mu}=\sum _f m_f\bar {f}f+
2m_z^2 Z^{\mu}Z_{\mu}+2m^2_w W^{\mu}W_{\mu}$. 
The couplings of $\tphi$ to SM fields are therefore suppressed
 by $\vphi$ which is in the Tev range. Since conformal invariance is 
broken explicitly by the mass of the SM fields, the effects of $\tphi$
on low energy phenomenology 
( i.e. much below the Kaluza-Klein excitations) can become 
appreciable only if the SM fields involved are quite heavy and
lie in the few hundred Gev range. In discussing the low energy
phenomenology of the radion we shall consider particularly those
processes that involve the couplings of $\tphi$ to the heavy SM
fields e.g. Z, W and t.
 
\vfill\eject

\centerline{\bf Loop induced virtual effects}

Let us first consider the effects of $\tphi$ through loop induced
radiative corrections which can be tested through high precision
tests. In this category we shall discuss the effect 
of $\tphi$ on anomalous $W^+W^-\gm$, $W^+W^-Z$ couplings and the 
renormalization of $Zt_R\bar{t}_R$ coupling. Besides we shall also
consider the radion contribution to the magnetic moment anomaly of
the muon. Unless stated otherwise we shall assume that $\vphi$= 1 Tev.

i) Anomalous magnetic moment of W boson: The anomalous
magnetic moment of W boson is given by the operator [4]
$$L_{WWV}=ie\k_v W^+_{\mu}W^-{\nu}V^{\mu\nu}.\eqno(2)$$

where $V^{\mu\nu}=\partial^{\mu}V^{\nu}-\partial^{\nu}V^{\mu}$ and 
V=A, Z. $\k_v$ is the anomalous magnetic moment of the W boson.
The value of $\k_v$ arising from SM interactions at one loop is of the 
order of $6\times 10^{-3}$. The radion contribution to $\k_v$ arises
from only one diagram. After pulling out one momentum corresponding
to the incoming $\gm$ or Z the remaining scalar integration can be evaluated
by naive power counting. For $\mphi \le m_w$ 
the integral gets most of its contribution from loop momenta of order 
$m_W$ and is given by $I(m_{\phi}^2, m_w^2)\approx {1\over 16\pi^2 m_w^2}
f({\mphi^2\over m_w^2})$ where $ f(\mu^2)=\int_0^1 x^2
{x^2-x+2+{\mu^2\over 2}(1-x)\over x^2+\mu^2(1-x)}$ and $\mu ={\mphi\over
m_w}$. For $\mphi\ll m_w$
the radion contribution to $\k_v$ is therefore given by 
$\k^{\tphi}_v\approx  {1\over 16\pi^2}{m_w^2\over \vphi^2}
 f({\mphi^2\over m_w^2})\approx
3.7\times10^{-4}$.
The anomalous magnetic moment of the W boson arising from radion
exchange is therefore roughly one order of magnitude smaller than 
that from SM processes.
The size of the  above radiative correction   however decreases inversely as 
$\vphi ^2$. Thus for $\vphi = $ 500 Gev the value of $\k^{\tphi}_v$
increases to $1.5\times 10^{-3}$.
However the dependence of $\k^{\tphi}_v$ on ${\mphi^2\over m_w^2}$ is 
almost flat over the entire range of the variable and therefore no 
useful bound on $\mphi$ can be obtained.

ii) Electric quadrupole moment of W boson: The electric quadrupole moment 
of the W boson is given by the operator [4]
$$L_{WWV}=ie{\lambda_v\over m_w^2} W^+_{\lambda\mu}W^{-\mu}_{\nu}
V^{\nu\lambda}.\eqno(3)$$. 

In SM $\l_v$ is zero at tree level. As in the case of $\k_v$ the radion
contribution to $\l_v$ arises from only one diagram. To evaluate
the electric quadrupole moment of the W boson arising from radion exchange
we need to extract three momenta
corresponding to the three external gauge bosons outside the loop
integral. The remaining integral can be shown to converge as 
${1\over m_w^4}g({\mphi^2\over m_w^2})$.
The radion contribution to $\l_v$ is therefore given by $\l_v\approx
{1\over 16\pi^2}{m_w^2\over \vphi^2}g({\mphi^2\over m_w^2})
\approx (1.6-3.2)\times 10^{-4}$ where we have taken $g(x)$ to lie between
1 and 2. At one loop order SM processes gives rise to a $\l_v$
that is of order $5\times 10^{-3}$. The modulus contribution to
$\l_v$ is therefore suppressed relative to the SM contribution
by one order of magnitude.

The sensitivity reach of different colliders for measuring $\k_v$
and $\l_v$ have been estimated by several working groups. These studies [5] 
indicate that LEPII with a design luminosity of 500 pb$^{-1}$ will be able 
to measure $\k_v$ and $\l_v$ with a precision of a few times $10^{-2}$.
On the other hand NLC 500 with a design luminosity of 10 fb$^{-1}$  will
be able to reach precision levels of about $10^{-3}$  which will enable
it to probe the radion contributions for $\vphi$ up to 500 Gev.
If $\vphi$ lies in the few Tev range then
 NLC 500 will not be able to probe the radion 
contribution to $\k_v$ and $\l_v$.
However we have seen that increasing 
the center of mass
energy from 200 Gev to 500 Gev enhances the sensitivity  reach
for measuring $\k_v$ and $\l_v$ by one
order of magnitude. Hence a multi Tev $e^+e^-$ collider with a very
high luminosity may be able to probe the small radion contribution 
to $\k_v$ and $\l_v$ for $\vphi$ = 1 Tev.

iii) Effect of $\tphi$ on the renormalization of $Zt\bar {t}$ couplings:
The radion contribution on the renormalization of $Zt\bar {t}$ couplings
is important because top quark is by far the heaviest known particle 
and has the strongest coupling to the radion. The stabilized modulus
of Golberger and Wise is expected to be lighter than 1 Tev and therefore
it would contribute to the renormalization of Zt$\bar{t}$ coupling
from $m_t$ to 1 Tev.
 There are two distinct
vertex correction diagrams due to $\tphi$ exchange. The dominant diagram
involves the exchange of $\tphi$ between the outgoing $t_R$ and 
$\bar {t}_R$. The vertex renormalization constant corresponding to this
diagram is given by $Z^{1}_v=1-{1\over 32\pi^2}\ln{q^2\over m_t^2}{g^t_L
\over g^t_R}({\mt\over \vphi })^2$ where q is the momentum of the 
incoming Z boson, $g^t_L={1\over 2}-{2\over 3}\sin^2\theta_w$
and $g^t_R=-{2\over 3}\sin^2\theta_w$.
In the second diagram Z goes into virtual $\tphi$ and $Z$ which then 
exchanges a top quark to produce a $t_R$, $\bar{t}_R$ pair in the final
state. This diagram involves a chirality flip through a mass insertion
on the internal top quark line. The renormalization constant corresponding
to this diagram does not involve a leading log term and therefore it
does not contribute to the renormalization of $Zt\bar{t}$ couplings.
The self energy correction of 
Z due to $\tphi$ produces only a mass renormalization of Z but no 
wavefunction renormalization. The wavefunction renormalization of
$t_R$ arising from $\tphi$ exchange is given by $Z_{t_R}=1+{1\over 32\pi^2}
({\mt\over \vphi})^2 \ln{q^2\over \mt^2}$. The effect of a light radion
on the renormalization of $g^t_R$ from $\mt^2$ to $q^2=(1000Gev)^2$ is 
therefore given by
$${[g^t_R(q)]_{RS}\over [g^t_R(q)]_{SM}}=1+(1-{g^t_L\over g^t_R})
({\mt\over \vphi})^2{1\over 32\pi^2}\ln {q^2\over \mt^2}.\eqno(4)$$

where we have assumed that
$g^t_R(\mt )_{RS}=g^t_R(\mt )_{SM}=g^t_R(\mt )_{expt}$.
The presence of an extra light field ($\tphi$) in the RS scenario  
causes the renormalization of $g^t_R$ to deviate from that in the SM
by .1\% for $\vphi $= 1 Tev. The splitting increases to .004\%
for $\vphi $=500 Gev.
 This effect is of the same magnitude as the radiative corrections
due to ordinary QED processes and it might be possible to detect this effect
at a multi Tev scale $e^+e^-$ collider. Note that the splitting between 
$[g^t_R(q)]_{RS}$ and $[g^t_R(q)]_{SM}$ increases logarithmically
with the high energy scale at which they are compared.

iv) The muon magnetic moment anomaly: Although the muon coupling to the 
radion is small the muon magnetic moment anomaly is an extremely well 
measured quantity. The present experimental value of $a_{\mu}$ is given
by [6]
$$a^{exp}_{\mu}=(116592.30\pm .8)\times 10^{-8}\eqno(5)$$

The extremely high precision with which $a_{\mu}$ can be measured
 compensates for the low radion coupling strength to the muon. The radion
contribution to the muon anomaly arises from only one diagram and
is given by

$$a_{\mu}^{\tphi}={1\over 4\pi^2}({m_{\mu}\over \vphi})^2\int_0^1 dx
{x^2(2-x)\over x^2+r(1-x)}.\eqno(6)$$
where $r=({m_{\tphi}\over m_{\mu}})^2$ is a free parameter. 
An upper bound on $a_{\mu}^{\tphi}$ can be obtained by setting r=0.
In this limit we get $a_{\mu}^{\tphi}\approx 4.4\times 10^{-10}$.
This value should be compared with the present experimental precision
of $10^{-8}$ in measuring $a_{\mu}$. The BNL experiment (E821)[7] hopes to 
lower the error the error down to $4\times 10^{-10}$. At that level
it will be able to probe the small radion contribution to $a_{\mu}$.
As ${\mphi^2\over m_{\mu}^2}$ becomes much greater than one the 
value of $a_{\mu}^{\tphi}$ decreases from the above estimate.
However for all $\mphi\le m_w$ the dependence of $a^{\tphi}_{\mu}$
 on ${\mphi}$ is almost flat.
So  even with the BNL precision neither a sharp nor a useful bound
can be put on $m_{\phi}$. It is worth noting that the bound from
muon anomaly experiments are relevant only for very light radion
($m_{\phi}\le m_{\mu}$). For such low radion mass the stabilization
of the modulus is very weak.

iv) Radiative mixing between the radion and the higgs:
Since the couplings of the radion and the higgs boson 
to SM particles are similar
in structure they can mix through loop corrections. The dominant
correction comes from the top loop because of a color factor and 
its large yukawa coupling. It can be shown that this contribution
is given by

$$\delta m^2_{\phi}\approx - {3N_c\over 4 \pi^2}{m^2_t\over v\vphi}
m^2_t \ln {m^2_t\over \mu^2}\eqno(7)$$
where $\mu $ is the renormalization scale. For $\mu =m_h=m_{\phi}
\approx 100 $ Gev the mixing angle is given by $\theta \approx
{\delta m^2_{\phi}\over m_h^2}\approx $-.11. This mixing between the 
radion and the higgs boson will shift the coupling of the physical
higgs boson to W and Z. For example the WWh coupling is now
given by

$$L_{wwh}= [gm_w + 2m_w {m_w\over \vphi}\theta ] WWh=gm_w[1-.026]WWh.
\eqno(8)$$.
 which implies a 2.6\% correction to the WWh coupling. The shift
in the Z coupling due to $\phi -h$ mixing is also of the same order.
Needless to say that probing this shift  is a challenging  
task but it should be reachable at NLC 500 with a high luminosity.

\centerline{\bf Direct production processes}

Let us now consider the effects arising from direct production of 
$\tphi$. The prominent among them is the radionstrahlung at an $e^+e^-$
collider. It involves the radiation of $\tphi$ from an on shell
(LEP I) or off shell (LEP II or NLC) Z boson. This process is similar
to higgsstrahlung which is used for higgs search at $e^+e^-$ collider.
At LEPII higgs boson is produced by the process $e^+e^-\rightarrow
Z^*\rightarrow h+Z$. It can be shown that the momentum of the outgoing
Z boson is given by $p_h^z={1\over 2\sqrt{s}}[s-(m_h-m_z)^2]^{1\over 2}
[s-(m_h+m_z)^2]^{1\over 2}$. Hence the momentum of the outgoing Z
boson for a 50 Gev higgs production at $\sqrt{s}$=160 Gev will be given
by $p_h^z$= 37 Gev. On the other hand for a 20 Gev radion emission its 
momentum will be given by $p^z_{\phi}$=51 Gev. So by imposing the cut 
$p_z > 45 $ Gev it should be possible to suppress the SM higgs production
for $m_h \ge$ 50 Gev while still allowing considerable amount of 
radion production with a lower mass.

The $\tphi ZZ$ coupling is suppressed relative to hZZ coupling by
a factor of $({\sin 2\theta_w\over e}{m_z\over \vphi})$. Hence for a given
$\sqrt{s}$ the production cross section for $\tphi$ will be comparable
to that of higgs(h) only if $m_{\tphi}$ is light but $m_h$ is close
to its kinematic limit so that its production is phase space suppressed.
We find that at LEPII for a cm energy of 189 Gev, 
$\sigma_{\tphi z}\approx$ .089 pb for $m_{\tphi}=20$ Gev and 
$\vphi $ =1 Tev, but 
$\sigma_{hz}=.078$ pb for $m_h=97$ Gev. For $m_{\tphi}=20$ Gev
and $m_h=97$ Gev the branching ratio for $\phi\rightarrow b\bar{b}$
and $h\rightarrow b\bar{b}$ are both roughly equal to one. Hence in this 
case the number of $q\bar{q}b\bar{b}$ events for example arising from
$e^+e^-\rightarrow hz$ will be roughly equal to that arising from
$e^+e^-\rightarrow \tphi z$. Let us consider the extreme scenario 
where the higgs is quite heavy so that it is kinematically inaccessible
at LEPII but $m_{\tphi}$ lies in the 20 Gev range. In this case one
can derive a lower bound on $m_{\tphi}$ from the condition $S\le 2\sqrt
{S+B}$ where is S is the signal and B is the background.
The L3 collaboration at LEPII have placed a lower limit of 96 Gev on the
higgs mass at 95\%  CL based on 176 pb$^{-1}$ of data collected at
$\sqrt{s}=189$ Gev [8]. If we make the somewhat unrealistic assumption that 
the cut efficiencies and background estimates are the same for the 
radion then this would imply a lower bound of about 20 Gev on the radion
mass. It should howver be noted that the optimal cuts that are necessary
to enhace the signal to background ratio for a 20 Gev
radion will be quite different  from  that for a 97 Gev higgs. Therefore
to obtain
a realistic bound on $m_{\tphi}$ from LEPII the different working
groups will have to retune their cuts appropriately so that they
 correspond to
light ($m_{\phi}\approx 20$ Gev) radion detection. 
Finally let us consider the implication of LEPI data on the radion mass.
At LEPI the higgs scalar has been searched based on $Z\rightarrow hZ^*$
with the $Z^*$ going into $l\bar{l}$.
 LEPI has placed a lower 
bound of about 60 Gev on $m_h$ by looking for this decay mode.
 The branching ratio for the above
 decay mode for $m_h $= 60 Gev is around $1.4\times 10^{-6}$. The smaller
radion coupling to the Z boson implies that the same branching ratio
will be attained for $Z\rightarrow {\tphi}l\bar {l}$ at $\mphi =35 $
Gev. We therefore think that the present collider data 
on direct higgs searches does not rule
out the existence of a light radion above 35 Gev.
We would like to point out that these values are indications
of the likely bounds on the radion mass
and should not be considered as realistic collider bounds.
More dedicated searches and detailed analysis are necessary to arrive at
realistic collider bounds.

\centerline{\bf Conclusions}

In this paper we have studied some low energy effects of a light
stabilized radion in the R-S scenario.  We have found that the radion 
contribution to the anomalous magnetic moment and electric quadrupole
moment of the W boson will fall beyond the projected sensitivity
reach of NLC 500 if $\vphi $=1 Tev.
 However the planned BNL experiment might be able to reach
a sensitivity level just adequate for probing the radion contribution to
$a_{\mu}$ if the radion mass is around 100 Mev. We have also shown that
 the higgs-radion mixing through loop corrections induces
a shift in the WWh coupling of the order of 2.6\%.
 The size of the radiative corrections arising from radion
 exchange however decreases inversely as $\vphi ^2$.
Therefore with decreasing $\vphi$ the radiative corrections
increase rapidly in size and approach the
search limits.
The dependence of the radiative corrections on the radion mass
however does not produce any useful bound on it.
 We have also shown that the presence of a light radion field
will cause the renormalization of $Zt_R\bar{t}_R$ coupling to deviate 
from that in the SM by about .1\%. Finally by comparing the radion 
production cross section at LEPII with that of the higgs boson we 
find that the lower bound of 96 Gev on the higgs 
mass suggests a  bound of about 20 Gev on the radion mass. The LEPI
bound of 60 Gev on $m_h$ however implies a slightly stronger bound
of 35 Gev  on $\mphi$.

\centerline{\bf Acknowledgement}

We would like to thank Dr. Sandip Trivedi, Dr. Sunanda Banerjee
and Dr. Anindya Datta
for several helpful discussions during the WHEPP6 working group activity.
S. Rakshit would like to thank the Council for Scientific
and Industrial Research, India
for support during this work.

\centerline{\bf References}

\item{1.} W.D. Goldberger and M. B. Wise: Phys. Rev. Lett 83, 4922
(1999); Phys. Rev. D 60, 107505 (1999).

\item{2.} L. Randall and R. Sundrum: Phys. Rev. Lett. 83, 3370 (1999);
Phys. Rev. Lett. 83, 4690 (1999).

\item{3.} W. D. Goldberger and M. B. Wise, hep-ph/9911457. C. Csaki,
M. Graesser, L. Randall and J. Terning, hep-ph/9911406.

\item{4.} K. Hagiwara, R. Peccei, D. Zeppenfield and K. Hikasa,
Nucl. Phys. B 282, 253 (1987).

\item{5.} Tests of alternative models at a 500 Gev NLC, ENSLAPP-A-365/92.
Published in ``$e^+e^-$ collisions at 500 Gev, the physics potential''
ed. P. Zerwas, DESY, 1992.

\item{6.} Caso et al., Particle Data group: Review of Particle Physics,
Euro. Phys. J. C3, 1 (1998).

\item{7.} By E821 BNL collaboration (B. C. Roberts et al.). Published in
the proceedings of XXVIII International Conference on High Energy Physics,
Warsaw, Poland, 1996, River Edge, NJ World Scientific, 1997, p.1035.

\item{8.} Search for the SM higgs boson in $e^+e^-$ interactions at 
$\sqrt{s}$= 189 Gev, L3 collaboration, CERN-EP/99-080.

\end